\documentclass[aps,prl,twocolumn,superscriptaddress]{revtex4}

\usepackage[dvips]{graphicx,color}

\input epsf.sty
\usepackage{graphicx}

\begin{document}

\preprint{\today}

%
%Title of paper
%

\title{Charge excitations associated with charge stripe order in the 214-type 
nickelate and superconducting cuprate}

\author{S. Wakimoto}
%\author{Shuichi Wakimoto}
%\email[Corresponding author: ]{wakimoto.shuichi@jaea.go.jp}
\affiliation{ Quantum Beam Science Directorate, Japan Atomic Energy Agency,
   Tokai, Ibaraki 319-1195, Japan }

\author{H. Kimura}
%\author{Hiroyuki Kimura}
\affiliation{ Institute of Multidisciplinary Research for Advanced Materials, 
   Tohoku University, Sendai 980-8577, Japan }

\author{K. Ishii}
%\author{Kenji Ishii}
\affiliation{ Synchrotron Radiation Research Unit, Japan Atomic Energy Agency,
   Hyogo 679-5148, Japan }

\author{K. Ikeuchi}
%\author{Kazuhiko Ikeuchi}
\affiliation{ Synchrotron Radiation Research Unit, Japan Atomic Energy Agency,
   Hyogo 679-5148, Japan }

\author{T. Adachi}
%\author{Tadashi Adachi}
\affiliation{ Department of Applied Physics, Tohoku University, 
   Sendai 980-8579, Japan }

\author{M. Fujita}
%\author{Masaki Fujita}
\affiliation{ Institute for Materials Research, Tohoku University, Katahira,
   Sendai 980-8577, Japan }

\author{K. Kakurai}
%\author{Kazuhisa Kakurai}
\affiliation{ Quantum Beam Science Directorate, Japan Atomic Energy Agency,
   Tokai, Ibaraki 319-1195, Japan }

\author{Y. Koike}
%\author{Yoji Koike}
\affiliation{ Department of Applied Physics, Tohoku University, 
   Sendai 980-8579, Japan }

\author{J. Mizuki}
%\author{Jun'ichiro Mizuki}
\affiliation{ Synchrotron Radiation Research Unit, Japan Atomic Energy Agency,
   Hyogo 679-5148, Japan }

\author{Y. Noda}
%\author{Yukio Noda}
\affiliation{ Institute of Multidisciplinary Research for Advanced Materials, 
   Tohoku University, Sendai 980-8577, Japan }

\author{A. H. Said}
%\author{Ayman H. Said}
\affiliation{ Advanced Photon Source, Argonne National Laboratory, 
   Argonne, Illinois 60439, USA }

\author{Y. Shvyd'ko}
%\author{Yuri Shvyd'ko}
\affiliation{ Advanced Photon Source, Argonne National Laboratory, 
   Argonne, Illinois 60439, USA }

\author{K. Yamada}
%\author{Kazuyoshi Yamada}
\affiliation{ Institute for Materials Research, Tohoku University, Katahira,
   Sendai 980-8577, Japan }
\affiliation{ Advanced Institute for Materials Research, Katahira,
   Sendai 980-8577, Japan }

\date{\today}

\begin{abstract}

Charge excitations were studied for stipe-ordered 214 compounds, 
La$_{5/3}$Sr$_{1/3}$NiO$_{4}$ and 1/8-doped La$_{2}$(Ba, Sr)$_{x}$CuO$_{4}$ 
using resonant inelastic x-ray scattering in hard x-ray regime.  We have 
observed charge excitations at the energy transfer of $\sim 1$~eV with the 
momentum transfer corresponding to the charge stripe spatial period both for 
the diagonal (nikelate) and parallel (cuprates) stripes.  These new excitations 
can be interpreted as a collective stripe excitation or charge excitonic mode 
to a stripe-related in-gap state.

\end{abstract}

\pacs{74.72.Dn, 74.25.Jb, 78.70.Ck}

\maketitle

There is accumulation of theoretical and experimental implications that the
inhomogeneous charge state, such as a charge stripe state, occurring as a
result of the carrier doping into the strongly correlated electron system is 
intimately related to the realization of  high temperature superconductivity 
in cuprates~\cite{Kive_rev03,Zaanen_rev01}.  
Understanding the dynamics of the charge stripe state is one of the basic 
problems.  To date, whereas the dynamics in spin sector has been studied 
extensively by neutron 
scattering~\cite{Birgeneau_JPSJ06,Hayden_Nature04,Tra_Nature04,Woo_PRB07}, the 
dynamics in charge sector is yet to be understood.

The 214 type nickelates and cuprates, family compounds of the firstly 
discovered high temperature superconductor La$_{2-x}$Ba$_{x}$CuO$_{4}$, exhibit 
charge stripe order when holes are doped.  The stripe order has been 
comprehensively studied by diffraction 
techniques~\cite{Tra_PRB96,Vigliante_PRB97,Tra_Nature95,Zimm_EPL98} and 
characterized by antiphase antiferromagnetic domains divided by one dimensional 
charge stripes where doped holes are confined.  Upon carrier doping, 
La$_{2-x}$Sr$_{x}$NiO$_{4}$ (LSNO) shows ordered state of ``diagonal'' stripes, 
where the charge stripes run along the diagonal direction of the NiO$_{2}$ 
square lattice~\cite{Tra_PRB96} (Fig. 1 a).  Hole-doped 214 cuprates 
La$_{2-x}$(Ba, or Sr)$_{x}$CuO$_{4}$ (LBCO or LSCO) show diagonal stripe order 
in the low doping insulating region~\cite{Waki_PRB99}, whereas in the 
superconducting dome the systems, particularly LBCO, show static order of 
``parallel'' stripes only in the vicinity of 
$x=0.125$~\cite{Suzuki_PRB98,Fujita_PRB04} where the superconductivity is 
suppressed~\cite{Mooden_PRB88}.  This is frequently quoted as 1/8 anomaly.  
Here the charge stripes are parallel to the Cu-O-Cu bond (Fig. 1 b).  
Consequently, relation between the dynamical fluctuation of parallel stripes 
and the superconductivity is of interest.

% Figure 1 %%%%%%%%%%%%%%%%%%%%%%%%%%%%
\begin{figure}
\includegraphics[width=3.2in]{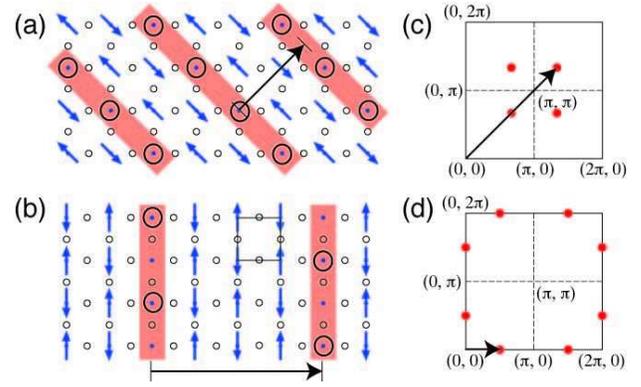}
\caption{(Color online) 
Diagonal charge stripes in (a) LSNO $x=1/3$ and parallel stripes in (b) LBCO 
$x=0.125$ and LSCO $x=0.12$ in the basal (Ni, or Cu)O$_2$ plane.  Stripes are 
marked by shaded belts.  (c) and (d) depict momentum space for the 
(Ni or Cu)O$_{2}$ square lattice in unit of $1/a$ ($a$ is the lattice constant 
of the square lattice).
Arrows connecting the stripes in (a) and (b) indicate spatial periods of stripes.
Arrows in (c) and (d) indicate characteristic momentum transfer ${\rm \bf q_s}$.
The period of $(3a/2, 3a/2)$ in (a) gives characteristic momentum transfer 
${\rm \bf q_s} = (2\pi/3, 2\pi/3)$, and the period of $(4a, 0)$ in (b) gives 
${\rm \bf q_s} =  (\pi/2, 0)$.  Circles in (c) and (d) indicate equivalent 
${\rm \bf q_s}$ points due to the stripe domains whose stripes run perpendicular 
with each other.
}
\end{figure}
%%%%%%%%%%%%%%%%%%%%%%%%%%%%%%%%%%%

Resonant inelastic x-ray scattering (RIXS) in hard x-ray regime, a recently 
developed experimental probe, have revealed various types of charge excitations 
in the strongly correlated electron systems, such as the charge excitation 
across the charge transfer (CT) 
gap~\cite{Hasan_Sci00,Kim_PRL02,Collart_PRL06,Ishii_PRL05}, intraband 
excitations~\cite{Ishii_PRL05}, molecular orbital excitations~\cite{Kim_PRB04_2}, 
$d-d$ excitations~\cite{Collart_PRL06}, collective charge 
excitations~\cite{Wray_PRB07}, and even magnons~\cite{Hill_PRL08}.  
The major advantage of RIXS is the possibility of studying charge excitations 
as a function of the momentum transfer {\bf q}.
The stripe spacing in the real space corresponds to a characteristic momentum 
transfer ${\rm \bf q_s}$ as shown in Figs. 1 c and  1 d.  RIXS is an ideal 
probe for detecting charge excitations associated with the stripe having a 
specific momentum transfer ${\rm \bf q_s}$.

In this Letter, we report the first observation of charge excitations arising 
from the charge stripe ordered state in 214 type nickelate, 
La$_{5/3}$Sr$_{1/3}$NiO$_{4}$, and 214 type superconducting cuprates 
La$_{1.875}$Ba$_{0.125}$CuO$_{4}$ and La$_{1.88}$Sr$_{0.12}$CuO$_{4}$ using the 
RIXS in hard x-ray regime.
Our measurements reveal the charge excitations with the momentum transfer 
corresponding to the charge stripe spatial period.  We observe this nontrivial 
feature both for the diagonal (nikelate) and parallel (cuprates) stripes.

Single crystals of LSNO $x=1/3$, LBCO $x=0.125$, LBCO $x=0.08$, and LSCO 
$x=0.12$ were prepared by the travelling solvent floating zone method.  The 
samples of LSNO $x=1/3$ and LBCO $x=0.125$ show robust charge stripe order 
below 180~K and 55~K, respectively.  LBCO $x=0.08$ shows no clear stripe order 
according to the neutron diffraction studies.  LSCO $x=0.12$ is expected to 
have stripe order below 30~K from neutron diffraction, but charge order has 
not been confirmed directly.  
Lattice constants for LSNO and L(B,S)CO at low temperatures give reciprocal 
lattice unit of $1.66$~\AA$^{-1}$ for the basal (Ni, or Cu)O$_{2}$ square 
lattices.  
The RIXS measurements of nickelate were performed at 10~K using the MERIX 
spectrometer installed at the Advanced Photon Source, beamline XOR-IXS 30-ID. 
RIXS measurements of cuprates were done at 8~K using the inelastic x-ray 
spectrometer at BL11XU at SPring-8.  Horizontal scattering geometry was utilized 
for all measurements, with the scattering plane parallel to the $(a, c)$ crystal 
plane.
Since the 214 compounds are two dimensional, we assign the $(0, 0, L)$ position 
as the $\Gamma$ point $(0, 0)$ of the basal plane and $(1, 0, L)$ as the next 
$\Gamma$ point $(2\pi, 0)$.  Here $L$ is chosen so that the scattering angle 
$2\theta$ is $\sim 90^{\circ}$.  In this configuration the scattered photon 
propagates parallel to the polarization of the initial photon, and thus, 
minimizes the elastic intensity.  This is crucial for observing the low energy 
excitations below $1.5$~eV reported in this Letter.

RIXS in the hard x-ray regime uses incident photons with the energy at the 
absorption K-edge of the transition metal element. Incident photons excite $1s$ 
core electrons into either $4p_\pi$ or $4p_\sigma$ orbitals depending on the 
sample geometry, and this intermediate state triggers various charge excitations.
In the present studies the incident photon energy was tuned to the energy of  
the $1s \rightarrow 4p_\pi$ transition, which is 8347 eV for nickelate and 
8993 eV for cuprates.  These energies are a few electron volts lower than the 
energy of the $1s \rightarrow 4p_\sigma$ transition.  
The instrumental energy resolution of the MERIX spectrometer at Ni K-edge is 
$150$~meV.  
This is achieved by using a Ge$(642)$ spherical diced analyzer, and a position 
sensitive microstrip detector placed on the Rowland circle of a $1$~m radius.  
The silicon microtrip detector with $125$~$\mu$m pitch is applied with the 
purpose of reducing the geometrical broadening of the spectral resolution 
function~\cite{Huorti_SR05}.  
The energy resolution of the Cu K-edge RIXS spectrometer at BL11XU is $400$~meV.
Bent Ge$(337)$ analyzer with $2$~m curvature radius and a point silicon detector 
is used.
$q$-resolutions of MERIX with the present configuration are 0.26~\AA$^{-1}$ and 
0.38~\AA$^{-1}$ along the $[\pi, 0]$ and $[0, \pi]$ directions, respectively.  
Those for BL11XU are 0.10~\AA$^{-1}$ and 0.15~\AA$^{-1}$, respectively.

% Figure 2 %%%%%%%%%%%%%%%%%%%%%%%%%%%%
\begin{figure}
\includegraphics[width=3.2in]{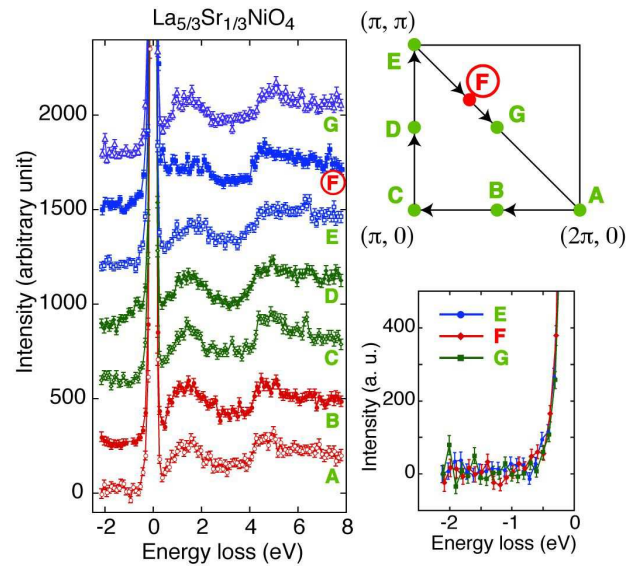}
\caption{(Color online) RIXS spectra for LSNO $x=1/3$ at several ${\rm \bf q}$ 
positions.  Each spectrum is shifted by 300 in intensity for clarity.  Measured 
${\rm \bf q}$ positions are indicated in the right top figure by circles 
labeled as A to G connected by allows, where the allows show the order of 
spectra from the bottom to the top.  The point, labeled F, is ${\rm \bf q_s}$.  
The right bottom figure shows comparison of elastic tails on the energy gain 
side of the RIXS spectra for positions at E, F, and G.}
\end{figure}
%%%%%%%%%%%%%%%%%%%%%%%%%%%%%%%%%%%

Figure 2 shows representative RIXS spectra for LSNO $x=1/3$ taken at the 
${\rm \bf q}$ positions indicated in the right top panel.  Every spectrum 
except that at ${\rm \bf q_s}=(4\pi/3, 2\pi/3)$ (labeled as F), contains mainly 
three peaks: an elastic peak at zero energy, and peaks at $\sim 1.5$~eV and  
$\sim 4.5$~eV.  The 4.5 eV feature is known to be the charge excitation across 
the CT gap (called a CT peak), which is also observed in the non-doped 
La$_2$NiO$_4$~\cite{Collart_PRL06}.  The $1.5$~eV feature can be attributed 
to the charge excitation from the valence band to the in-gap band which is 
known to appear when holes are doped into nickelates by optical measurements~\cite{Ido_PRB91,Katsufuji_PRB96,Homes_PRB03}.  
There is a dip between the elastic peak and the in-gap peak implying a gap in 
charge excitation, consistent with the fact that the nickelate is an insulator.  
However, to our surprise, the dip appears to be filled at 
${\rm \bf q}={\rm \bf q_s}$.
We note that the elastic peak at ${\rm \bf q}={\rm \bf q_s}$ is twice as 
strong as those at the other ${\rm \bf q}$ positions due to the charge stripe 
order~\cite{elastic}.  One might suspect that the additional intensity comes 
from the tail of the stronger elastic peak.  However, as clearly seen in the 
right bottom panel of Fig. 2, the intensity of
the elastic peak tail on the energy gain side at ${\rm \bf q}={\rm \bf q_s}$ 
is same as for other ${\rm \bf q}$ positions.  This rules out such explanation.

% Figure 3 %%%%%%%%%%%%%%%%%%%%%%%%%%%%
\begin{figure}
\includegraphics[width=3.2in]{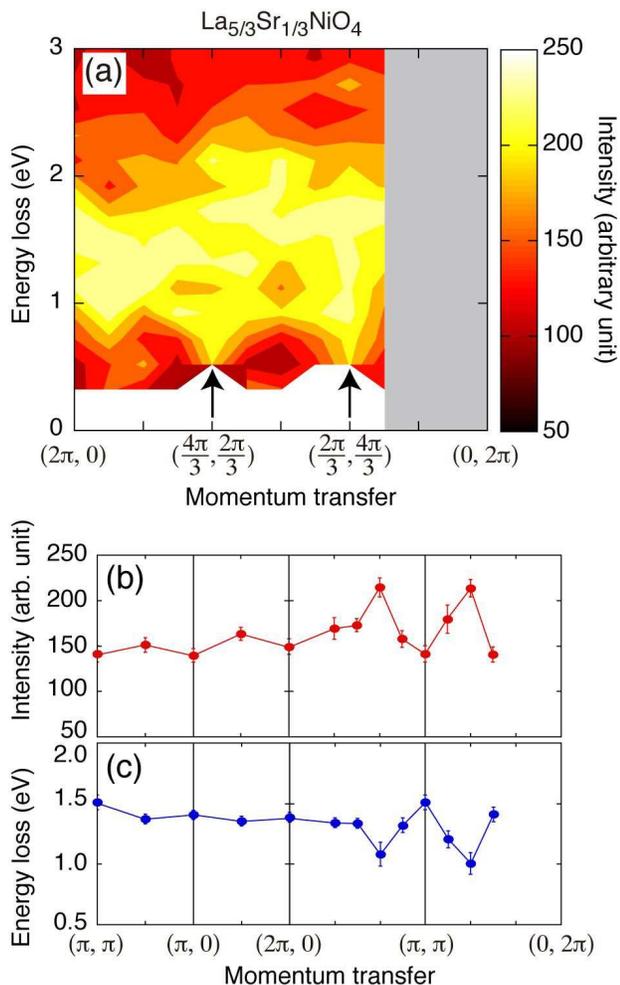}
\caption{(Color online) (a) A contour plot of the RIXS intensity between 
$(2\pi, 0)$ and $(0, 2\pi)$.  The white area has intensity more than 250 due 
to elastic peak or filled dip at ${\rm \bf q_s}$ positions.  The shaded 
rectangular area has not been measured.   
(b) ${\rm \bf q}$ dependence of the RIXS intensity averaged in the energy 
region between 0.5 and 0.9 eV (upper panel) and the position of the in-gap 
peak (lower panel).}
\end{figure}
%%%%%%%%%%%%%%%%%%%%%%%%%%%%%%%%%%%

We have measured RIXS spectra at several ${\rm \bf q}$-positions between 
$(2\pi, 0)$ and $(0, 2\pi)$ to test if the filled gap
%additional intensity 
is unique at ${\rm \bf q_s}$.  The RIXS intensities are mapped in Fig. 3 a as 
a contour plot.  It is clearly demonstrated that the dip between the elastic 
and the in-gap peaks is indeed absent at both ${\rm \bf q_s}$ positions, 
$(4\pi/3, 2\pi/3)$ and $(2\pi/3, 4\pi/3)$.
Correspondingly, the RIXS signal averaged in the energy region between $0.5$ 
and $0.9$~eV has maxima at ${\rm \bf q_s}$ (Fig. 3 b).
Figure 3 c shows in-gap peak positions obtained by 
%fitting the RIXS spectra using Lorentzian function.
fitting the RIXS spectra above 0.4 eV to a function with two Lorentzians and 
sloped background.  
It is shown that the in-gap feature is primarily independent of ${\rm \bf q}$ 
but the anomaly appears at ${\rm \bf q_s}$ due to the filled dip.

% Figure 4 %%%%%%%%%%%%%%%%%%%%%%%%%%%%
\begin{figure}
\includegraphics[width=3.2in]{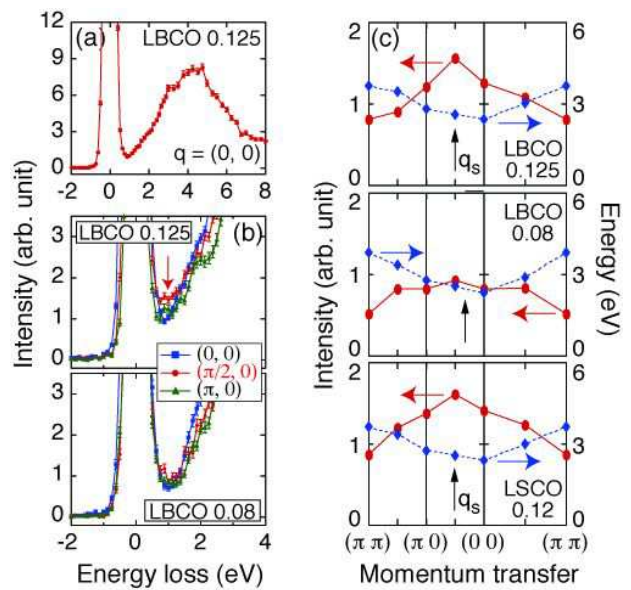}
\caption{(Color online) %RIXS spectra for the 214 cuprates.  
(a) Representative RIXS spectrum of LBCO $x=0.125$ at zone center.  
(b) Close-ups of RIXS spectra around $1$~eV of LBCO $x=0.125$ (upper) and 
LBCO $x=0.08$ (lower) measured at $(0, 0)$, $(\pi/2, 0)$, and $(\pi, 0)$.  
RIXS intensities are normalized to the maximum of the CT peak at $\sim 4$~eV.  
The arrow in (b) highlights anomalous enhancement of RIXS intensity around 
1~eV at ${\rm \bf q_s} = (\pi/2, 0)$ observed in LBCO $x=0.125$.  (c) Momentum 
transfer dependence of the RIXS signal averaged in the energy region between 
1 and 1.25 eV (circles) and the energy where the RIXS intensity reaches the 
half-maximum of the CT peak (diamonds).
Arrows in (c) indicate ${\rm \bf q_s}$ positions.  Although the LBCO $x=0.08$ 
sample does not have robust stripe order, it may have very small signal of 
static order with ${\rm \bf q_s} = 2\pi(2x, 0)$ as observed in underdoped 
LSCO~\cite{Hiraka}.  The arrow in the middle panel of (c) indicates this 
${\rm \bf q_s}= (0.32\pi, 0)$ for LBCO $x=0.08$ as a reference.}
\end{figure}
%%%%%%%%%%%%%%%%%%%%%%%%%%%%%%%%%%%

As a next step we will compare the LSNO results with those of cuprates whose 
representative RIXS spectrum is shown in Fig. 4 a.  All spectra we observed 
contain two peaks; the elastic peak and the CT peak at $\sim 4$~eV.  In the 
case of cuprates, the in-gap peaks are not observed clearly.  It was earlier 
reported that the continuum-like intensity near 1 eV increases with 
doping~\cite{Kim_PRB04} due to the increase of free (conductive) 
holes~\cite{Uchida_PRB91}.  A similar effect is seen in Fig. 4 b, showing 
the spectra of LBCO $x=0.125$ and $0.08$.
The parallel stripe order with a $4a$ spacing, which realizes in LBCO $x=0.125$ 
and LSCO $x=0.12$, gives ${\rm \bf q_s}$ at $(\pi/2, 0)$ (Fig. 1 c).  
Remarkably, we have found an anomalous enhancement of the RIXS intensity near 
1 eV at ${\rm \bf q_s} = (\pi/2, 0)$ in the stripe ordered samples.  This is 
more clearly demonstrated in Fig. 4 c, where the RIXS intensity averaged in 
the energy range between 1 and 1.25 eV (circles) peaks at ${\rm \bf q_s}$ in 
LBCO $x=0.125$ and LSCO $x=0.12$.  On the contrary, it is flat in the 
non-stripe ordered LBCO $x=0.08$.
Diamonds in Fig. 4 c indicate the energy position where the CT peak reaches a 
half of its maximal intensity.  This quantity increases, that is, the CT peak 
shifts to the higher energy, as ${\rm \bf q}$ changes from $(0, 0)$ to 
$(\pi, 0)$ for all compounds.  Upon the same change of ${\rm \bf q}$, the 
elastic peak intensity decreases~\cite{elastic}.  These facts evidence that 
the enhancement of the RIXS intensity at $\sim 1$~eV relates to an intrinsic 
property of the system, and cannot be explained by the overlap of the tails 
of the elastic and CT peaks.

The above presented data on nickelate and cuprates demonstrate that the 
charge stripe state gives the additional RIXS intensity near $1$~eV at 
the ${\rm \bf q_s}$ positions regardless of the stripe geometry.  Here 
we discuss possible origins of the discovered excitation.
The calculations presented in Ref.~[\onlinecite{Kaneshita_PRL02}] predicted 
that the parallel stripes are accompanied by collective charge excitations 
with the momentum transfer ${\rm \bf q_s}$.  Qualitatively similar 
excitations are expected for the diagonal stripes.  RIXS can detect 
such excitations.  Therefore, it is reasonable to attribute the observed 
additional intensity at ${\rm \bf q}={\rm \bf q_s}$ in the RIXS spectra 
of Figs. 2 and 4 to the collective stripe excitations.
The instrumental resolutions in energy and momentum transfer of current 
setups are not good enough to see detailed structure of dispersion.  
Nevertheless the presented results demonstrate the potential of the RIXS 
technique for observing collective charge stripe excitations.

Another plausible explanation is that the anomaly at ${\rm \bf q_s}$ is 
the nature of charge exciton to the in-gap state; that is, the excitonic 
mode has anomalous softening at ${\rm \bf q_s}$.
An exciton is the bound state between an excited electron to the in-gap 
band and a left hole in the valence band, and hence the exciton spectrum 
must be always gapped.  In this case, the anomalous softening of the exciton 
mode gives smaller gap at ${\rm \bf q_s}$ which we estimate to be less than 
340 meV for nickelate and less than 800 meV for cuprates from the 
instrumental resolution.

It is not an easy task to distinguish between the above two possibilities.  The
collective charge excitation should be possible to be detected by non-resonant 
IXS whereas a creation of the charge exciton requires the resonant process.  
Thus performing the non-resonant IXS is one way to proceed.  However, intensity 
of the collective charge excitation might be extremely small. 
Also, theoretical calculations of in-gap band are desirable to examine the 
excitonic possibility.  The origin of the in-gap state has been interpreted 
in many ways~\cite{Katsufuji_PRB96,Homes_PRB03,Tsutsui_PRB99}.   
The present observation evidences the connection between the in-gap state and 
the stripe.  It is important to test if the softening of excitonic mode is 
reproducible by taking the charge stripe structure into account whatever the 
origin of the in-gap state is.

In summary, we have performed RIXS measurements in hard x-ray regime of the 
stripe-ordered 214 type nickelate and superconducting cuprates.  We have 
observed for the first time charge excitations at the characteristic momentum 
transfer ${\rm \bf q_s}$ related to the spatial period of the stripes.  They 
can be interpreted as a collective stripe excitation or charge excitonic 
mode to a stripe-related in-gap state.

% Acknowledgements

Authors thank E. Kaneshita, K. Machida, and K. Nakajima for invaluable 
discussions, and Peter Siddons (BNL) for building the microstrip detector 
of MERIX.
This work is partially supported by a Grant-In-Aid from the Japanese Ministry 
of Education, Culture, Sports, Science and Technology.
Use of the Advanced Photon Source was supported by the U. S. Department of 
Energy, Office of Science, Office of Basic Energy Sciences, under Contract 
No. DE-AC02-06CH11357.

%\newpage

\end{document}